\documentclass[twocolumn]{aastex631}
\usepackage{multirow}
\usepackage{makecell}

\newcommand{\astfootnote}[1]{%
\let\oldthefootnote=\thefootnote%
\setcounter{footnote}{0}%
\renewcommand{\thefootnote}{\fnsymbol{footnote}}%
\footnote{#1}%
\let\thefootnote=\oldthefootnote%
}

\begin{document}

\title{GRB~221009A: Observations with LST-1 of CTAO and implications for structured jets in long gamma-ray bursts}   

\author{K.~Abe}
\affiliation{Department of Physics, Tokai University, 4-1-1, Kita-Kaname, Hiratsuka, Kanagawa 259-1292, Japan}
\author[0000-0001-7250-3596]{S.~Abe}
\affiliation{Institute for Cosmic Ray Research, University of Tokyo, 5-1-5, Kashiwa-no-ha, Kashiwa, Chiba 277-8582, Japan}
\author[0009-0005-5239-7905]{A.~Abhishek}
\affiliation{INFN and Università degli Studi di Siena, Dipartimento di Scienze Fisiche, della Terra e dell'Ambiente (DSFTA), Sezione di Fisica, Via Roma 56, 53100 Siena, Italy}
\author[0000-0002-6606-2816]{F.~Acero}
\affiliation{Université Paris-Saclay, Université Paris Cité, CEA, CNRS, AIM, F-91191 Gif-sur-Yvette Cedex, France}
\affiliation{FSLAC IRL 2009, CNRS/IAC, La Laguna, Tenerife, Spain}
\author[0000-0001-8816-4920]{A.~Aguasca-Cabot}
\affiliation{Departament de Física Quàntica i Astrofísica, Institut de Ciències del Cosmos, Universitat de Barcelona, IEEC-UB, Martí i Franquès, 1, 08028, Barcelona, Spain}
\author[0000-0002-3777-6182]{I.~Agudo}
\affiliation{Instituto de Astrofísica de Andalucía-CSIC, Glorieta de la Astronomía s/n, 18008, Granada, Spain}
\author{C.~Alispach}
\affiliation{Department of Astronomy, University of Geneva, Chemin d'Ecogia 16, CH-1290 Versoix, Switzerland}
\author{D.~Ambrosino}
\affiliation{INFN Sezione di Napoli, Via Cintia, ed. G, 80126 Napoli, Italy}
\author{F.~Ambrosino}
\affiliation{INAF - Osservatorio Astronomico di Roma, Via di Frascati 33, 00040, Monteporzio Catone, Italy}
\author[0000-0002-5037-9034]{L.~A.~Antonelli}
\affiliation{INAF - Osservatorio Astronomico di Roma, Via di Frascati 33, 00040, Monteporzio Catone, Italy}
% ...
\collaboration{331}{(the CTAO-LST collaboration)}
\correspondingauthor{A. Aguasca-Cabot, S. Inoue, \mbox{M. Seglar~Arroyo}, and K. Terauchi}

\email{lst-contact@cta-observatory.org}



\begin{abstract}
GRB\,221009A is the brightest gamma-ray burst (GRB) observed to date.
Extensive observations of its afterglow emission across the electromagnetic spectrum were performed, providing the first strong evidence of a jet with a nontrivial angular structure in a long GRB.
We carried out an extensive observation campaign in very-high-energy (VHE) gamma rays with the first Large-Sized Telescope of the future Cherenkov Telescope Array Observatory, starting on 2022 October 10, about 1 day after the burst. 
A dedicated analysis of the GRB~221009A data is performed to account for the different moonlight conditions under which data were recorded. We find an excess of gamma-like events with a statistical significance of 4.1$\sigma$ during the observations taken 1.33 days after the burst, followed by background-compatible results for the later days.
The results are compared with various models of afterglows from structured jets that are consistent with the published multiwavelength data but entail significant quantitative and qualitative differences in the VHE emission after 1 day.
We disfavor models that imply VHE flux at 1 day considerably above $10^{-11}\,\rm{erg\, cm^{-2}\, s^{-1}}$.
Our late-time VHE observations can help disentangle the degeneracy among the models and provide valuable new insight into the structure of GRB jets.

\end{abstract}


\section{Introduction}\label{sect:introduction}

    Gamma-ray bursts (GRBs) are brief, intense flashes of gamma rays peaking in the MeV band, detected at an average rate of $\sim$1 day$^{-1}$, randomly distributed in the sky \citep{Meegan1992Spatial}. 
    The initial, prompt phase of their emission exhibits irregular variability and typically lasts seconds to minutes \citep{fishman1994first}.
    This is followed by the afterglow phase, where emission across the electromagnetic spectrum decays more gradually, over timescales of hours up to months \citep{vanParadijs2000}.
    Based on the duration and spectra of the prompt emission, they are classified as either short- or long-duration GRBs \citep{Kouveliotou93}.
    Long GRBs are known to typically originate from the core collapse of some massive stars \citep{Woosley&Bloom2006}, while short GRBs are widely thought to be triggered by the coalescence of binary compact objects \citep{Berger2014, Margutti2021}. 
    For either GRB class, the event is believed to generate collimated jets of plasma with ultrarelativistic bulk velocities, within which the prompt emission is produced and observed as a GRB when the jet is oriented close to our line of sight \citep{Rees1994}.
    The afterglow emission from the radio band up to the GeV band is robustly interpreted as synchrotron radiation by electrons accelerated in a blast wave, triggered by the interaction of the jet with the ambient medium \citep{Meszaros1997, Sari1998, Piran2010}. Inverse Compton radiation by the same electrons can induce afterglows at even higher photon energies \citep{Fan2008}. For reviews on GRBs, see \citet{Meszaros2002}, \citet{Piran2004}, \citet{Gehrels2012}, and \citet{Kumar2015}.
    
    Gamma-ray emission at very high energies (VHEs; $E>100\,\textrm{GeV}$) from GRBs had been long expected \citep{Meszaros1994, Zhang2001, Kakuwa2012, Inoue_2013, Nava_2018} but was not observationally verified until recently, with the detection of VHE gamma-ray emission from four different GRBs with the Major Atmospheric Gamma Imaging Cherenkov (MAGIC) and High Energy Stereoscopic System (H.E.S.S.) telescope facilities: GRB\,190114C \citep{MAGIC190114C}, GRB\,180720B \citep{abdalla2019very}, GRB\,190829A \citep{HESS190829A}, and GRB\,201216C \citep{Abe_2023_GRB201216C}. These detections confirmed that at least some long GRBs emit VHE gamma rays during the afterglow phase \citep{Nava_2021, Miceli2022, Noda2022}. For short GRBs, a $\sim$3$\sigma$ hint was reported by MAGIC for GRB~160821B \citep{acciari2021magic}.
   
    On 2022 October 9 at 13:16:59.99 UTC, hereafter $T_0$, the Gamma-ray Burst Monitor (GBM) on board the Fermi Gamma-Ray Space Telescope detected an extremely bright burst at 0.01--1\,MeV lasting hundreds of seconds \citep{GBM2022GCN,Lesage_2023}\footnote{A bright, line-like emission feature of unknown origin with temporal evolution in both energy (from $\sim$12 to $\sim$6 MeV) and luminosity (from $\sim$$10^{50}$ to $2\times 10^{49}$\,erg\,s$^{-1}$)  was identified in the Fermi-GBM data \citep{Ravasio:2024ete}.}. 
    The Swift Burst Alert Telescope (BAT) reported the detection of a very bright transient at 15--150\,keV
    at 14:10:17 UTC at the coordinates $(\mathrm{R.A.}, \mathrm{decl.})(\mathrm{J}2000)=(19^\mathrm{h} 13^\mathrm{m} 3\fs43,19\arcdeg46\arcmin 16.3\arcsec)$ \citep{Swift2022GCN_xraytrans, Williams2023swift}, triggering follow-up observations by other instruments.
    From the detections by Fermi-GBM and Swift-BAT
    coincident in time and localization, the source was recognized as an extremely bright long GRB, labeled GRB\,221009A \citep[][]{Swift2022GCN_GRB}.
    Other instruments on satellites such as AGILE-GRID \citep{tavani2023agile}, Insight-HXMT \citep{Insight-HXMT_2022Atel}, and GECAM-C \citep{GECAM-C_2022GCN} also detected the event.
    From optical spectroscopic follow-up observations, the redshift of the source was determined to be $z=0.1505$ \citep{GTCGCN2022}.
    In high-energy (HE) gamma rays ($E>100$\,MeV), the Fermi Large Area Telescope (LAT) reported extremely bright emission \citep{FermiLAT2022GCN}, the bulk of which started $\sim$200 s after the GBM trigger pulse and manifested rapid variability in flux and spectra for $\sim$200 s afterward
   \citep{axelsson2024FermiLAT}.         
    Due to the brightness of the event, Fermi-LAT suffered from a strong pileup at early times. During the prompt phase that lasted for more than 600\,s, a photon of 99.3\,GeV was detected at $T_0+240$\,s, while a photon of 400\,GeV was detected at $T_0+\sim$9 hr in the afterglow phase. These are the highest-energy photons seen by Fermi-LAT from a GRB during each phase \citep{FermiLATrefined2022GCN,axelsson2024FermiLAT}. The Large High Altitude Air Shower Observatory (LHAASO) was observing the region of the sky that included GRB~221009A during the prompt and afterglow phases and reported the detection of VHE gamma rays from the GRB by the WCDA detector between 200\,GeV and 7\,TeV at more than 250$\sigma$ significance \citep{LHAASO2022GCN,LHAASO2023WCDA}.  
    LHAASO also reported the detection of the GRB with the KM2A detector from $\simeq$3 to $\simeq$13\,TeV during the period $T_0+[230,900]$\,s \citep{LHAASO2023KM2A}. The High-Altitude Water Cherenkov (HAWC) gamma-ray observatory reported observations starting at $T_0~+\sim8$\,hr, providing a preliminary differential flux upper limit (UL) at 1\,TeV \citep{HAWC2022GCN}.
    Rapid follow-up observations of GRB~221009A by imaging atmospheric Cherenkov telescopes (IACTs) were prevented by the brightness of the full Moon on October 9. H.E.S.S. observations started at $T_0+53$\,hr and ULs were reported between 650\,GeV and 10\,TeV
    \citep{HESS2023}.

    Follow-up observations were also conducted at all wavelengths spanning the radio, optical, and X-ray bands, resulting in the most extensive multiwavelength (MWL) coverage of a long GRB to date \citep{Kann2023, Laskar2023, OConnor2023, Williams2023swift, radio_Giarratana2024, Rhodes2024}.
    This led to some unique inferences regarding the underlying physical processes.
    First, the temporal and spectral properties of the VHE gamma-ray and X-ray emission seen up to a few thousand seconds showed that they likely originate from an afterglow due to a narrow jet with an opening angle of $\sim$0$\rlap{.}\arcdeg$6  \citep{HXMTandGECAMC2023, LHAASO2023WCDA}.
    On the other hand, the temporal behavior of the radio-to-X-ray emission at later times cannot be explained by such a narrow jet and requires a separate emission region.
    This is most plausibly identified with a wider outer jet surrounding the narrower inner jet
    \citep{Gill2023, Sato2023_BOAT, Ren2024, Zheng2024, Zhang2025}.
    Such structured jets, for which basic jet parameters like the kinetic energy and bulk velocity depend on the angle from the jet axis in a non trivial way \citep{Meszaros1998, Rossi2002, Zhang2002}, are naturally expected in realistic models of jet formation and propagation in GRBs \citep{Morsony2007, Mizuta2013, Gottlieb2021}.
    Notwithstanding some indications in previous GRBs \citep{Salafia2022, Sato2023_2jet}, GRB~221009A represents the first long GRB with strong evidence for a structured jet.
    It offers a unique opportunity to probe the physics of jet formation and propagation in long GRBs, which is still not well understood \citep{Kumar2015}.
    However, afterglow models accounting for such structured jets are more complicated compared to standard, simpler afterglow models, and effectively constraining them requires comprehensive MWL observations.
    In particular, more VHE gamma-ray data at late times are highly desirable, in addition to the ULs obtained by H.E.S.S. and HAWC.

    The first Large-Sized Telescope (LST-1) of the Cherenkov Telescope Array Observatory (CTAO), inaugurated at the Observatorio del Roque de los Muchachos in 2018 October, has been taking science data since 2019 November. This is the first of the four LSTs \citep{LSTProject2023PoS} that are part of the 
    northern array of CTAO. LST-1 started observing GRB~221009A at 1.33\,days after the burst and continued for more than 20 days. 
    It constitutes the largest GRB campaign conducted by LST-1 to date, with deep coverage of the late afterglow phase. The analysis of the first 2 days of data required meticulous treatment of the night sky background (NSB), as the observations were acquired under moonlight conditions. 

    In this work, we present the results for GRB~221009A obtained during the follow-up campaign with LST-1.
    This includes the earliest observations of GRB~221009A by an IACT in a period not covered by other VHE facilities.
    We contextualize these results, compare them with theoretical models of VHE afterglow emission from structured jets, and address the physical implications.  
    
    The Letter is organized as follows. Section~\ref{sect:observations} describes LST-1 and the observing conditions under which the data were obtained. The details of the data analysis are given in Section~\ref{sect:data_analysis}. The results of the LST-1 analysis are presented in Section~\ref{sect:results}. 
    We compare the obtained data with theoretical models and discuss the physical implications in
    Section~\ref{sect:discussion}. Finally, we provide our conclusions in Section~\ref{sect:conclusions}.

\section{Observations}\label{sect:observations}

    CTAO is the new-generation ground-based facility for VHE gamma-ray astronomy, currently under construction\footnote{\url{https://www.ctao.org/}} \citep{CTA_concept_2013}. When completed, its sensitivity at 1\,TeV will be an order of magnitude better than the previous generation of IACTs \citep{Acharyya2019_MCCTA}. Three types of telescopes with different designs constitute each CTAO array. The different telescope design is aimed to cover different portions of the full CTAO energy range between 20\,GeV and 300\,TeV. LSTs, with a \mbox{23\,m} diameter mirror dish, are the largest IACTs in CTAO. Designed to cover the low-energy band, LSTs are suited for the detection of transient sources at tens to hundreds of GeV 
    \citep{CTA2019book,Nozaki_2023_BLLac,Abe_2025_RSOph}. First, the 
    large reflective surface of about 400\,m$^{2}$ enables the detection of faint Cherenkov flashes from electromagnetic cascades of gamma rays at energies down to 20\,GeV \citep{Abe_2023_LSTPerformance}. 
    Second, the telescope is equipped with a 4$\rlap{.}\arcdeg$5 field-of-view (FoV) camera that allows coverage of broad sky regions, beneficial in following up alerts of transient sources with localization uncertainties of a few degrees. Third, the light design of the telescope structure allows for fast slewing between coordinates, up to 180\arcdeg\ in about 20\,s \citep[see LST parameter details in][]{LSTstatus_2015ICRC}. LST-1 is capable of automatically reacting in real time to transient alerts, relying on the Transient Handler \citep{CTALSTProject:2021lxr}. External triggers are received and processed to initiate an automatic telescope response. The alerts are provided by the General Coordinates Network (formerly known as GRB Coordinates Network) 
    via the event-streaming platform Kafka. Such observations are complemented by late-time campaigns that are planned and initiated offline. 
    
    The LST-1 camera is composed of 1855 high quantum efficiency photomultiplier tubes (PMTs) converting the Cherenkov light into electrical signals, which are subsequently recorded by a fast readout system.    
    \mbox{LST-1} is optimized to operate under \textit{astronomical} darkness and absence of the Moon (hereafter \textit{dark}). If PMTs are exposed to bright environments, they experience accelerated aging and a significant gain reduction. Therefore, around full Moon periods, observations are halted due to the high NSB from the Moon. However, observations in moonlight conditions are feasible. These observations, hereafter referred to as \textit{moonlight} observations, increase the duty cycle of IACTs by $\sim$30\% with respect to the $\sim$1500 hr\,yr$^{-1}$ of data in \textit{dark} conditions \citep{AHNEN201729,ARCHAMBAULT201734,OHM2023168442}. Increasing the duty cycle is relevant for all source types but is particularly critical for fast-evolving transient sources. Yet it comes at the cost of compromised telescope performance and higher systematic uncertainties on the estimated spectrum. Moonlight observations can be conducted with reduced high voltage (HV) to reduce the operational gain of the PMTs across all the NSB levels encountered when the Moon is above the horizon.

    \begin{table}
        \caption{
        Observations with \mbox{LST-1} of GRB~221009A in 2022 October.
        }
        \begin{center}
        \label{tab:obs_table}
        \begin{tabular}{lccc}       
        \hline \hline
        \noalign{\smallskip}
        Start Date &  $T-T_0$ & Time after & Zenith Angle \\
        & & Data Selection & Range \\        
        $\rm{(MJD)}$ & (days) & (hr) & (deg)\\
        \noalign{\smallskip}
        \hline
        \noalign{\smallskip}
         59862.88$^{\rm a}$ &  1.33 & 1.75 & 31--54\\
         59864.89$^{\rm a}$ &  3.33 & 1.42 & 34--52\\
         59867.85~\, &  6.30 & 0.80 & 25--52\\
         59868.88~\, &  7.32 & 2.35 & 34--65\\
         59869.85~\, &  8.30 & 2.41 & 28--60\\
         59875.86~\, &  14.30 & 2.01 & 34--61\\
         59877.89~\, &  16.33 & 1.18 & 45--59\\
         59878.87~\, &  17.32 & 1.42 & 42--58\\
        \noalign{\smallskip}
        \hline
        \end{tabular}
        \end{center}

        \tablecomments{For each observation day, the starting date, starting time offset with respect to the burst trigger \citep[$T_0$;][]{GBM2022GCN}, the observation time after the data selection, and the zenith angle range of the observations are shown.\\
        $^{\rm a}$ Data taken under bright moonlight conditions.}
        
    \end{table}

    The night after the detection of GRB~221009A, on 2022 October 9, no operations of LST-1 were possible due to the presence of the full Moon. 
    However, due to the exceptional nature of this event, observations were resumed with reduced HV on 2022 October 10 ($T_0+1.33$\,days) and continued on 2022 October 12 ($T_0+3.33$\,days), with camera problems preventing observations on 2022 October 11 (see Table~\ref{tab:obs_table}). A total of 3.17\,hr of data were acquired under bright moonlight conditions. The observation campaign continued until the end of 2022 November, extending for two Moon periods.  In this work, we focus on the observations in the \textit{moonlight} and \textit{dark} conditions of 2022 October. A total of 10.17\,hr of good-quality observations were obtained in dark conditions in 2022 October. This is summarized in Table~\ref{tab:obs_table}.
    
    LST-1 observations were performed in \textit{wobble} mode \citep{FOMIN1994137}, where the telescope points to regions offset from the source coordinates. Four pointing directions were chosen, each offset by 0$\rlap{.}\arcdeg$4 from the GRB position reported by Swift-BAT \citep{Swift2022GCN_xraytrans}.  Starting from the pointing with a positive offset in R.A. only, this is followed by those rotated by 90\arcdeg, 180\arcdeg, and 270\arcdeg\ around the GRB position. The observation time for each pointing was 20 minutes.

\section{Data analysis}\label{sect:data_analysis}
    
    The analysis of the data is divided according to the specific observing conditions during their acquisition. 
    We consider a moonlight-adapted analysis (see Section~\ref{sect:moon_analysis}) to process the first two observation nights (see Table~\ref{tab:obs_table}) and standard dark analysis to process the rest of the data (see Section~\ref{sect:dark_analysis}).
    Both analyses are performed with the dedicated software analysis pipeline \verb|cta-lstchain v0.10.5| \citep{ruben_lopez_coto_2023_10245571}. The two independent analysis chains in \verb|cta-lstchain|, i.e., the source-independent and the source-dependent analysis, are used in this work \citep{Abe_2023_LSTPerformance}. The former is the standard analysis scheme for CTAO and selected as the reference analysis here, while the latter is used to cross-check the consistency and robustness of the results. This comparison is presented in Appendix~\ref{sect:srcdep-analysis}.

\subsection{Analysis of observations in dark conditions}
\label{sect:dark_analysis}

    The LST-1 performance under dark conditions has been studied in \citet{Abe_2023_LSTPerformance}. A similar analysis is used for the GRB~221009A observations in dark conditions (see Table~\ref{tab:obs_table}). This approach considers a single calibration per observation obtained from calibration events (flat-field and pedestal events) taken in a dedicated observation run on the same night. The signals in the waveforms are integrated with the \verb|LocalPeakWindowSum| algorithm of \verb|ctapipe|  \citep[the prototype low-level data analysis package for CTAO;][]{ctapipe-icrc-2023}. Subsequently, we apply the tail-cut method for image cleaning, utilizing an increased picture threshold condition based on the pixel noise level. Additionally, a time-coincident condition and dynamic cleaning are used \citep{Abe_2023_LSTPerformance}.

    The energy, incident direction, and {\it gammaness\footnote{Gammaness is a score obtained with a random forest classifier; it ranges from 0 (very gamma-unlike image) to 1 (very gamma-like image).}} of the events are estimated through random forest algorithms, which are trained with Monte Carlo (MC) gamma rays and protons \citep{Abe_2023_LSTPerformance}.
       
    We need to adjust the NSB in the MC simulations to the specific NSB level in the FoV of GRB~221009A. Due to the proximity of GRB~221009A to the Galactic plane ($b\sim4\rlap{.}\arcdeg3$), the NSB level is similar to that for a Galactic source like the Crab Nebula. The extra NSB for the analysis of GRB~221009A is accounted for through the addition of random Poissonian noise on the camera images before the image-cleaning stage.

    The selection criteria for gamma-like events are optimized using data from the Crab Nebula, which were collected in 2022 under conditions similar to those of the GRB~221009A observations. We discard dim events with a total charge (intensity) below 50 photoelectrons (p.e.). Global cuts in the gammaness parameter and $\theta$ parameter\footnote{The angular separation between the reconstructed event and expected source position.} are used to assess the statistical significance of the detection. Energy-dependent, efficiency-based cuts on the gammaness and $\theta$ parameters are used for the spectral analysis. Three sets of gammaness cuts (50\%, 70\%, and 90\%) are tested, while the efficiency cut on the $\theta$ parameter is kept at 70\%. The Crab Nebula spectra obtained with these cuts are consistent with those in the literature. Among the three sets of cuts, the gammaness efficiency cut at 90\% provides the most stable results throughout the energy range studied and is selected for the analysis of GRB~221009A observations.
    
    The energy threshold of the analysis is defined as the peak energy position of the MC gamma-ray rates, weighted to the assumed spectral index of the source, after the event selection. 
    We consider an observed spectral index of $\Gamma_\mathrm{obs}=-3$ to compute the energy threshold, assuming that the spectral index at the time of the LST-1 observations is similar to that measured by LHAASO in the energy range between $\sim$0.2 and $\sim$7\,TeV at $T_0+[900,2000]\,\mathrm{s}$ \citep{LHAASO2023WCDA}. The energy threshold under this assumption for the dark data ranges from $\sim$20\,GeV at zenith angle 25\arcdeg\ to $\sim$200\,GeV at 65\arcdeg.
    
    We perform a spectral analysis using the analysis tools in the software package \texttt{gammapy v1.0} \citep{gammapyAA2023,gammapy_v1.0}. The joint-likelihood fitting method is used, in which three control regions (OFF) are employed to estimate the background. 
    For a given telescope pointing direction, these OFF regions are selected to be centered around it at the same angular distance to the GRB position and rotated by 90\arcdeg, 180\arcdeg, and 270\arcdeg\ from the GRB.     
    Only events above the energy threshold are used during the fitting process.

\subsection{Analysis of observations in moonlight conditions}
\label{sect:moon_analysis}

    The sensitive PMTs that are used to detect the faint Cherenkov light in LST-1 are affected by high-NSB conditions. Scattering of the moonlight in the atmosphere increases the anode current in the PMTs of the whole camera. Observations taken under moonlight conditions become noisy due to spurious NSB triggers, which produce larger noise fluctuations and more after-pulse signals in the waveforms as the NSB increases. As a consequence, the pulse timing and signal reconstruction are affected, worsening the precision in the event reconstruction and gamma/hadron separation compared to observations in dark conditions.   
    A dedicated data analysis is needed to ensure the best telescope performance under conditions with different trigger and camera settings (e.g., reduced HV) for moonlight observations. The following modifications to the standard analysis chain (see Section~\ref{sect:dark_analysis}) are made.

    The camera calibrations are adjusted to account for fast changes in the observing conditions: interleaved calibration events during the data taking are acquired to perform a continuous correction of the initial calibration parameters. In addition, we consider the algorithm \verb|NeighborPeakWindowSum| of \verb|ctapipe| to integrate the pulse of the waveform. \verb|NeighborPeakWindowSum| sums the signal over a window centered around the peak position, which is determined from the averaged waveform of the triggered pixel and its adjacent pixels. In particular, we considered a window size of seven waveform samples\footnote{One waveform sample is $\simeq$1\,ns.}, starting three samples before the estimated peak position. Concerning image cleaning, we employ the tail-cut method with the time-coincident condition and dynamic cleaning. The image-cleaning levels are increased by a factor of $\sim$2.5 compared with the values applied to the dark data. These values are determined by limiting the fraction of images from interleaved pedestal events that pass the image cleaning to less than 4\%.

    During the GRB observations in moonlight conditions, the NSB level increased during the data taking as the Moon rose. The NSB level between two consecutive 20 minutes observations is high enough to require a per-observation analysis,
    where the NSB on the MC simulations and image-cleaning levels is consequently adjusted. 
    The MC events are simulated with noisier waveforms to match the level of observed NSB in the data. Subsequently, we fine-tune the match between MC and real data on a per-observation basis at the camera image level by adding random Poissonian noise.

    We select gamma-like events and produce the final analysis products following the same procedure as for the analysis in dark conditions. For the Moon-adapted analysis, the event selection cuts are optimized with Crab Nebula observations in NSB conditions similar to that for GRB~221009A with reduced HV. These data were obtained on 2022 November 5 in the same zenith range as the GRB~221009A observations, during which the conditions evolved rapidly in a similar way. 
    A cut in intensity below 200\,p.e. is applied to remove dim events. 
    For the spectral analysis, we use energy-dependent, efficiency-based cuts of 50\% and 70\% for the gammaness and $\theta$ parameters, respectively. In this case, the tightest gammaness cut from the tested set is selected because the alternative cuts (70\% and 90\% in gammaness), while also yielding results consistent with the Crab Nebula spectrum reported in the literature, are more sensitive to variations in the lower energy bound of the fitting range.

    The tighter image cleaning with increasing NSB and different telescope pointing directions as a function of zenith angle affects the energy threshold of the analyses. In particular, for October 10, the energy threshold, assuming an observed spectral index of $\Gamma_\mathrm{obs}=-3$, increases from $\sim$100\,GeV during the first observation to 300\,GeV during the last.

\section{Results}\label{sect:results}

\begin{figure}[t]
        \centering
        \includegraphics[width=\hsize]{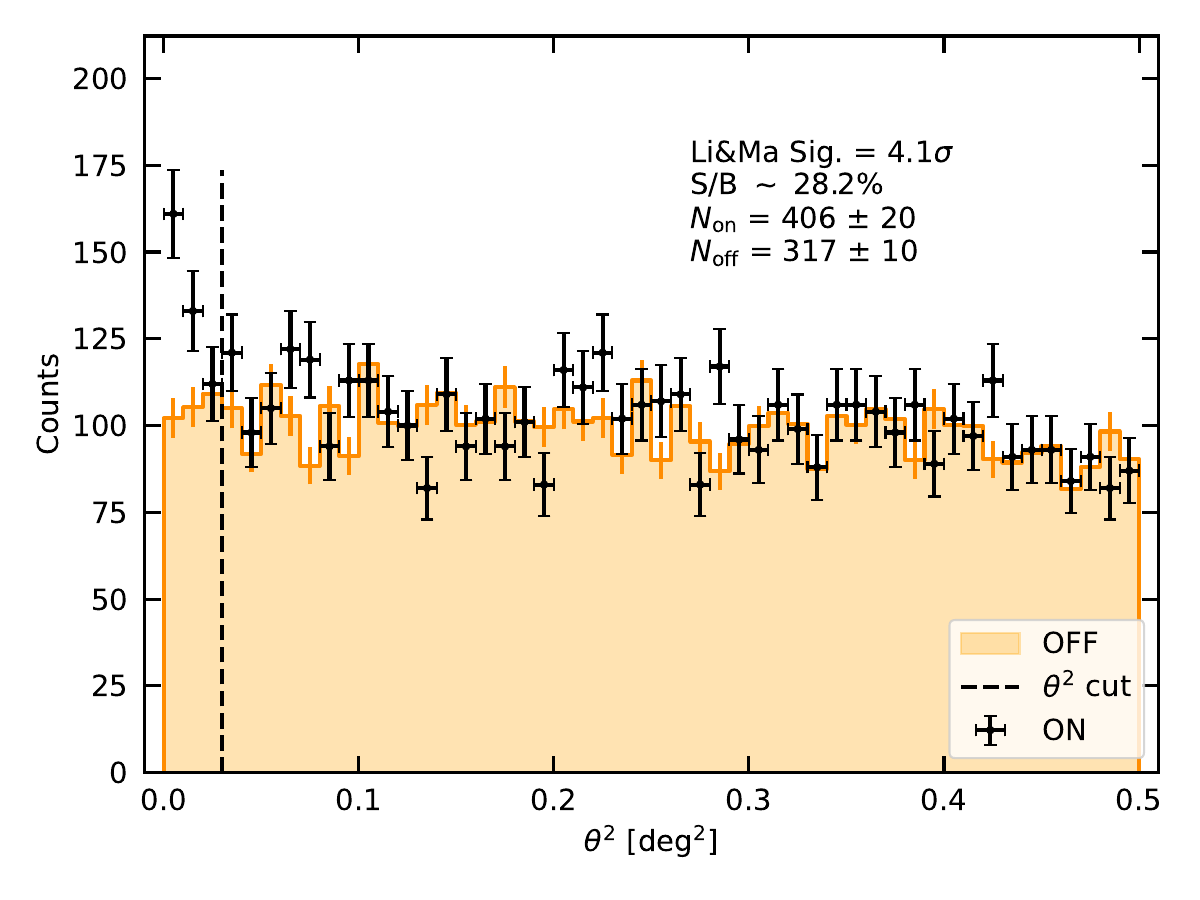}
        \caption{$\theta^2$ plot for observations at $T_0+1.33$\,days. The $\theta^2$ distributions centered at the GRB~221009A position (ON) and the mean background estimation from the three reflected regions (OFF) are displayed as black points and dark orange error bars, respectively. The vertical dashed line indicates the angular size used to compute the detection statistical significance (Li$\&$Ma Sig.) and signal-to-background ratio (S/B). The vertical error bars correspond to 1$\sigma$ statistical errors.}
        \label{fig:theta2plot_Oct10}
    \end{figure}

    The reconstructed squared angular distributions of gamma-like events centered on the GRB~221009A coordinates and the average background regions for the first day of observation ($T_0+1.33\,$days) are shown in Figure~\ref{fig:theta2plot_Oct10}. We obtain an excess 
    at 4.1$\sigma$ statistical significance \citep[using equation 17 of][]{LiMa83}. For the second day of observations at $T_0+3.33\,$days, the statistical significance is 0.8$\sigma$. As no signal was observed at $T_0+3.33\,$days, guided by the gradual power-law temporal decay observed at other wavelengths, we stack all the dark time data, which allows us to obtain better constraints on the spectral energy distribution (SED) of GRB~221009A. No significant excess is observed using all the dark observations together ($-0.4\sigma$ using data within $T_0+[6.30,17.32]\,$days). Results using the source-dependent analysis chain are consistent with that reported above for the source-independent analysis: at $T_0+1.33\,$days, the excess reaches a statistical significance of 4.6$\sigma$, while the significance is compatible with the background for the data sets at later times (see Appendix~\ref{sect:srcdep-analysis}).

    A power-law intrinsic spectrum is assumed for the putative emission from GRB~221009A. The expected attenuation due to the extragalactic background light (EBL) is accounted for using the \citet{Dominguez2011} model for redshift $z=0.1505$. We note that for such values of $z$, the choice of the EBL model is not critical.
    Given the limited significance of the excess,
    the intrinsic spectral index before EBL attenuation is fixed to $\Gamma=-2$ during the fitting process and the computation of the SED and light curve. This index is similar to that seen by LHAASO at much earlier epochs in the period $T_0+[900,2000]\,\mathrm{s}$ at energies between $\sim$0.2 and $\sim$7\, TeV \citep{LHAASO2023WCDA}. We also checked the case of assuming $\Gamma=-3$, and obtain comparable results.
    The range of $\Gamma$ we consider covers the value of $\Gamma \sim -2.5$ determined by Fermi-LAT at energies and times that overlap with the LST-1 observations \citep{axelsson2024FermiLAT}.
    ULs are computed at a 95\% confidence level when the test statistic $(\rm TS)$ is below 4. Error uncertainties correspond to 1$\sigma$ statistical errors.

\begin{figure}
        \centering
        \includegraphics[width=\hsize]{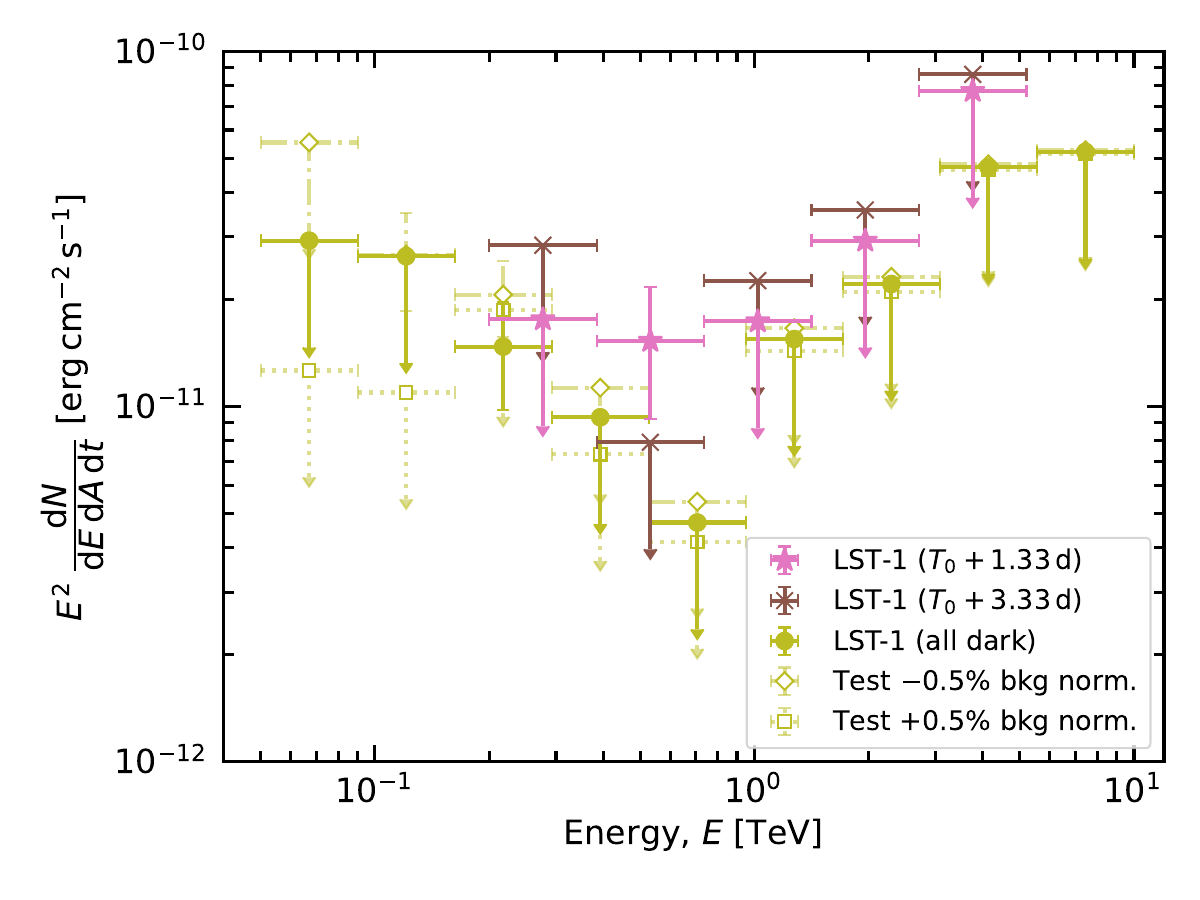}
        \caption{Intrinsic SED of GRB~221009A corrected for EBL attenuation on 2022 October 10 ($T_0+1.33\,$days; pink), 2022 October 12 ($T_0+3.33\,$days; brown) and between 2022 October 15 and 27 (all dark, $T_0+[6.30,17.32]\,$days; olive), respectively. For the latter SED, the open diamond and square olive markers show the effect of increasing and reducing by $0.5\%$ the normalization of the background, respectively. The vertical error bars correspond to 1$\sigma$ statistical errors, and ULs are computed at the 95\% confidence level.
        }
        \label{fig:SED-GRB221009A-LST1}
    \end{figure}

    Figure~\ref{fig:SED-GRB221009A-LST1} shows the SEDs for the three periods using the moonlight ($T_0+1.33$\,days and $T_0+3.33$\,days) and dark ($T_0+[6.30,17.32]\,$days) observations. The lower energy bound is 200\,GeV and 50\,GeV for the moonlight and dark analyses, respectively.
    We can constrain the EBL-corrected SED points to be below a few $10^{-11}\,\rm{erg\, cm^{-2}\, s^{-1}}$ at $E<1$\,TeV, with the most constraining ULs at several hundred GeV. For the first observation day, we obtain an SED point with a local $\rm TS=6.9$ in the energy bin between 0.38 and 0.74\,TeV.
    We obtain compatible SED results with the source-dependent analysis (see Appendix~\ref{sect:srcdep-analysis}), for which a significant SED point ($\rm TS=9.0$) is also obtained in the same energy bin for the observations at $T_0+1.33$\,days, while ULs constrain the emission at a similar differential flux level. 
    We note that the SED for the source-dependent analysis at $T_0+1.33$\,days is shifted toward higher flux values/ULs,  
    compared to the source-independent SED across the studied energy range. On the contrary, this shift is not visible for the $T_0+3.33$\,day and $T_0+[6.30,17.32]\,$day data (see Appendix~\ref{sect:srcdep-analysis}). The presence of this shift only at $T_0+1.33$\,days may be caused by enhanced systematic uncertainties due to the high-NSB conditions. Yet overall, no significant difference is observed between the two analyses for any of the periods.

    The effect of varying the background normalization by $\pm 0.5\%$ is shown in Figure~\ref{fig:SED-GRB221009A-LST1} to evaluate possible systematic errors in the background estimation. 
    A $\pm0.5\%$ relative difference in events between the control OFF regions and the mean OFF events are seen for the dark observations at the lowest energies ($E<200$\,GeV) where the number of events is large, $\mathcal{O}(10^5\textendash 10^6)$.    
    The modification of the background normalization by $\pm0.5$\% corresponds to a $\sim$60\% relative difference in the estimated SED ULs at the lowest energies for $T_0+[6.30,17.32]\,$days. 
    As pointed out in \citet{Abe_2023_LSTPerformance}, the monoscopic configuration of LST-1 entails modest background suppression close to the threshold of the telescope. On the contrary, the tighter cuts and higher-energy fit range used for the moonlight observations reduce the number of events to $\mathcal{O}(10^3)$, making the systematic uncertainty associated with the background normalization irrelevant for this data set. Applying the background normalization test on the estimated SEDs at $T_0+1.33$\,days and $T_0+3.33$\,days results in small changes (less than 3\%).

    The integral energy flux is computed keeping $\Gamma=-2$, the same value assumed in the SED derivation. The energy flux is computed between 0.3 and 5\,TeV for a clear comparison with data from other VHE experiments. Note that this depends on the choice of $\Gamma$ since the higher energies are more affected by EBL attenuation. If $\Gamma=-3$ is adopted, the energy flux is reduced by about a factor of 2. Additionally, the integration interval at high energies is loosely constrained compared to the lowest energies, where most of the observed excess lies. 
    
\begin{figure*}
        \centering
        \includegraphics[width=\hsize]{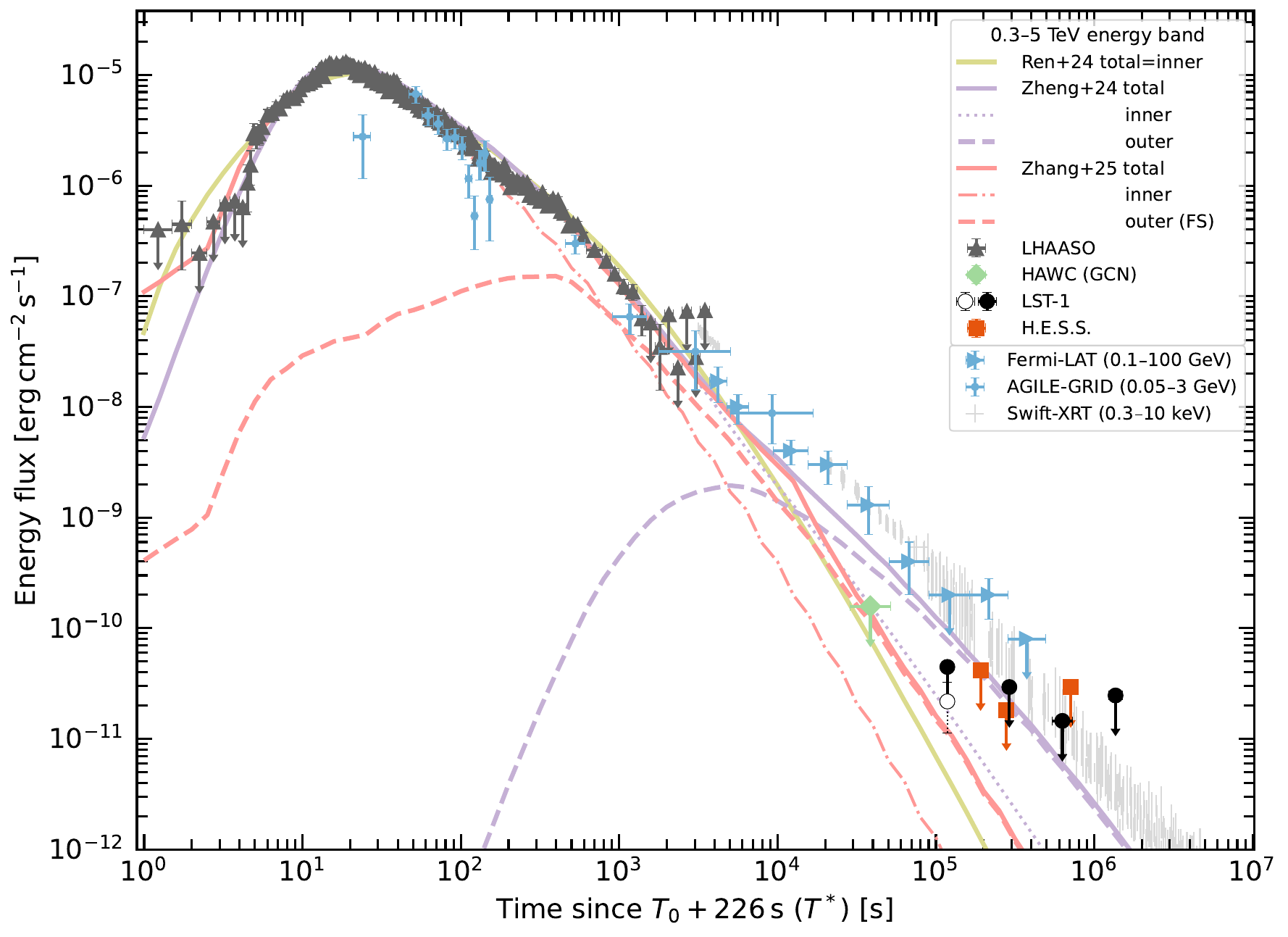}
        \caption{MWL intrinsic light curve of GRB~221009A corrected for EBL attenuation
        vs. time since the burst trigger ($T_0$) shifted by $+226$\,s (see text). The energy fluxes for the 0.3--5\,TeV band with LHAASO \citep[dark gray vertical triangles; ][]{LHAASO2023KM2A}, HAWC (green diamond; adapted from \citealt{HAWC2022GCN}; see text), LST-1 (black filled and open circles for ULs and the energy flux point, respectively; this work), and H.E.S.S. (orange squares; adapted from \citealt{HESS2023}; see text) are compared with the best-fit emission models from \citet{Ren2024}, \citet{Zheng2024}, and \citet{Zhang2025} in pale gold, purple, and red respectively.
        For the latter two models, the contributions from the inner and outer jet regions are also shown separately.
        In addition, the light curve in HE gamma rays with AGILE-GRID \citep[light blue dots; ][]{tavani2023agile} and Fermi-LAT \citep[light blue horizontal triangles; Extended Table 3 of][]{axelsson2024FermiLAT} as well as in X-rays with Swift-XRT \citep[light gray; ][]{Williams2023swift} are displayed. The vertical error bars correspond to 1$\sigma$ statistical errors, and ULs are computed at the 95\% confidence level.}
        \label{fig:eflux-data-grb221009A}
    \end{figure*}

    The energy flux is estimated in four different time periods, two for the moonlight observations ($T_0+1.33$\,days and $T_0+3.33$\,days) and two for the dark observations. The first subset of the dark data is recorded during intervals closer
    to the burst trigger ($T_0+[6.30,8.30]\,$days), while the second is for later intervals
    ($T_0+[14.30,17.32]\,$days).
    This is motivated by the wide time window of the dark observations and the lack of good-quality data for several days in a row (see Table \ref{tab:obs_table}). These results are shown in Figure~\ref{fig:eflux-data-grb221009A}, starting at about $10^5$\,s after the burst trigger.    

    At $T_0+1.33$\,days, both the energy flux and UL are shown, given the putative signal on this day ($\rm TS=4.6$ in the energy range of 0.3--5\,TeV). The ULs at $E=[0.3,5]$\,TeV constrain the EBL-corrected energy flux at the level of a few $10^{-11}\,\rm{erg\, cm^{-2}\, s^{-1}}$. 
    The light curve derived with the source-dependent analysis in Appendix~\ref{sect:srcdep-analysis} is consistent with that in Figure~\ref{fig:eflux-data-grb221009A}. However, at $T_0+1.33$\,days, the energy flux between 0.3 and 5\,TeV ($\rm TS=14.3$) for the source-dependent analysis is about 2 times higher than that for the source-independent analysis, differing by $1.5\sigma$ including the errors. This may be within the systematic errors that can be comparable to the statistical ones.

    The energy flux of GRB~221009A measured by different instruments is also shown in Figure~\ref{fig:eflux-data-grb221009A}. The light curve spans over 20 days with the reference time at $T^*=T_0+226$\,s, when LHAASO reveals the onset of the afterglow at VHE \citep[0.3--5\,TeV;][]{LHAASO2023WCDA}. Coincident with LHAASO, Fermi-LAT and AGILE-GRID detect GRB~221009A in the HE band \citep{tavani2023agile, axelsson2024FermiLAT}. In particular, extended emission up to few 10$^{5}$\,s is detected with Fermi-LAT (0.1--100\,GeV), coincident in time with part of the \mbox{LST-1} observations. In Figure~\ref{fig:eflux-data-grb221009A}, we include for reference only the HE light curve obtained with AGILE-GRID within 1\,ks after $T_0$. The corresponding Fermi-LAT light curve is excluded from this time interval to avoid overcrowding the figure. After several hours, a preliminary differential flux UL at 1\,TeV \citep{HAWC2022GCN} is reported by HAWC, shown in Figure~\ref{fig:eflux-data-grb221009A} for 0.3--5\,TeV by correcting for EBL attenuation with the \citet{Dominguez2011} model and assuming $\Gamma=-2$. HAWC bridges the early VHE observations by LHAASO and those by LST-1 at $T_0+1.33\,$days, the first by an IACT. Monitoring by IACTs continued during the following days, leading to several ULs with H.E.S.S. and LST-1. The original H.E.S.S. ULs at $E=[0.65,10]\,$TeV are recomputed for 0.3--5\,TeV using the spectral index value assumed in \citet{HESS2023}, which constrains the emission at a level similar to the LST-1 ULs. 
    These gamma-ray observations are summarized in Figure~\ref{fig:eflux-data-grb221009A}, where we also plot the Swift-XRT energy flux (0.3--10\,keV) for reference.

    In summary, a positive deviation in counts from a background-only hypothesis is obtained at 4.1$\sigma$ in the region of GRB~221009A with LST-1 at $T_0+1.33$\,days. Afterward, we measure no significant excess in nonconsecutive days at $T_0+[3.33,17.32]$\,days.  The SEDs with LST-1 probe the afterglow emission at the lowest energies of the VHE band, not previously studied with good sensitivity for GRB~221009A. In particular, the best differential sensitivity with LST-1 is obtained at a few hundred GeV \citep{Abe_2023_LSTPerformance}, where the effect of EBL attenuation is small.
    Our data provide deep constraints on the energy flux at the level of a few $10^{-11}\,\rm{erg\, cm^{-2}\, s^{-1}}$ at 0.3--5\,TeV
    after $T_0+1.33$\,days. The LST-1 observations on October 10 are the closest to $T_0$ obtained by an IACT, filling the gap between the HAWC UL (about 1 day before) and H.E.S.S. observations (about 1 day after).

\section{Discussion}\label{sect:discussion}

    The results obtained from observations of GRB~221009A with LST-1 described above are compared with selected theoretical models for the MWL emission of this burst, and the physical implications are addressed below.

    The VHE light curves of GRB~221009A observed by LHAASO in different energy bands are consistent with broken power laws, implying that most of the emission originates from the afterglow.
    A key finding is the achromatic (i.e. energy-independent) break in the light curve at $\sim T^*+670$\,s.
    This is most plausibly interpreted as a \textit{jet break}, caused by the decrease in emissivity when the opening angle of the relativistically beamed radiation from the decelerating blast wave becomes wider than that of the emitting jet plasma \citep{Rhoads1999, Sari1999}\footnote{For alternative interpretations, see \citet{Foffano2024} and \citet{Khangulyan2024}.}.
    The jet break time constrains the half-opening angle of the jet to be $\sim$0$\rlap{.}\arcdeg$6, much narrower than that inferred for most previously known GRBs \citep{LHAASO2023WCDA}.
    Coverage at other wavelengths contemporaneous with the LHAASO observations is sparse, but X-rays were measured at some early epochs by HXMT and GECAM-C with light curves similar to LHAASO, showing that the emission can originate from the same narrow jet \citep{HXMTandGECAMC2023}. If this narrow jet is the only emitting region, light curves at all other wavelengths after the break time are expected to be relatively steep, similar to that at VHE seen by LHAASO.
    However, the observed HE gamma-ray, X-ray, and optical light curves at $T \gtrsim T^*+$1000\,s reveal decay slopes that are considerably shallower,
    strongly indicating that an emission region separate from the narrow jet is necessary. The most likely such region is a wider outer jet surrounding the narrower inner jet. In general, physical properties of the jet such as the initial kinetic energy $E_{\rm kin,0}$ and bulk Lorentz factor $\Gamma_\mathrm{b,0}$ can be distributed as nontrivial functions of angle $\theta$ from the jet axis.
    GRB afterglow models assuming such jet configurations as initial conditions are referred to as ``structured jet'' models \citep{Meszaros1998, Salafia2022}, as opposed to the standard, simpler assumption employed in most earlier studies of ``top-hat'' jets, where $E_{\rm  kin,0}$ and $\Gamma_\mathrm{b,0}$ are distributed uniformly up to a certain angle.
    Structured jets are physically more realistic, as such configurations can arise when the jet forms \citep{Zhang2024jet} and are also robustly expected when the jet propagates through the progenitor star and interacts with stellar material \citep{Morsony2007, Mizuta2013, Gottlieb2021}.
    The afterglow of GRB~170817A, associated with a neutron star merger detected in gravitational waves, provided the first clear evidence for a structured jet in a GRB, albeit for an atypical short GRB \citep{Mooley2018}.
    For long GRBs, some previous studies suggested that the available data can be modeled better by structured jet models than simpler models \citep{Salafia2022, Sato2023_2jet}, but the conclusions were not definitive.
    GRB~221009A provides the strongest case to date that a structured jet is indispensable to explain the MWL afterglow of a long GRB \citep{Gill2023, OConnor2023, Sato2023_BOAT, Ren2024, Zheng2024, Zhang2025}.

    The MWL afterglow of GRB~221009A offers unique prospects for probing the jet structure in a long GRB that may be difficult to achieve otherwise.
    However, as the functional forms for $E_{\rm kin,0}(\theta)$ or $\Gamma_\mathrm{b,0}(\theta)$ are not known a priori, structured jet afterglow models necessarily entail a large number of parameters, often more than 20, compared to simpler afterglow models usually characterized by eight parameters.
    Even for the extensive MWL data obtained for GRB~221009A, this poses a challenge.

    This is illustrated by comparing three structured jet afterglow models in the literature that describe reasonably well the published MWL data of GRB~221009A: \citet{Ren2024}, \citet{Zheng2024}, and \citet{Zhang2025}.
    All three models assume the density profile of the circumburst medium $n_{\rm CBM}(R)$ to be constant with radius $R$ in the inner parts and to decline as $R^{-2}$ in the outer parts.
    Also common to all three are an inner jet component that is uniform but narrow, as required to account for the observed data before a few kiloseconds after $T^*$.
    At these epochs, the emission is attributed mainly to the forward shock of this inner jet, with X-rays and VHE gamma rays dominated by synchrotron and synchrotron self-Compton (SSC) emission, respectively, and HE gamma rays bridging these two components.
    The models also have outer jet components with assumed functional forms for $E_{\rm kin,0}(\theta)$ and $\Gamma_\mathrm{b,0}(\theta)$ that are somewhat different between the models, though not to a significant extent.
    Synchrotron emission from the forward shock of this outer jet is primarily responsible for the optical-to-X-ray emission starting from a few kiloseconds after $T^*$.
    For each model, the authors determined the best-fit values for the parameter set.

    Figure \ref{fig:eflux-data-grb221009A} compares the light curves at 0.3--5\,TeV for the available data including our LST-1 results with the models presented in \citet{Ren2024}, \citet{Zheng2024}, and \citet{Zhang2025}. 
    As a reference point for the power radiated in VHE gamma rays, we also show only the data for the X-ray and HE light curves. Corresponding model curves, as well as optical and radio data vs models, are shown in the papers above and not repeated here.
    Interestingly, although all models provide broadly acceptable fits to the published data set including LHAASO, the VHE emission of the models diverges at late times, differing by more than an order of magnitude in energy flux between them after $\sim T^*+1$\,days.
    Thus, in conjunction with the published MWL data, our LST-1 results provide valuable additional constraints on these structured jet models.

    First, conservatively considering our data at $T_0+1.33\,$days to be a UL, we disfavor the model by \citet{Zheng2024}, whose VHE flux around this time due to SSC from the outer jet is the highest among the three and which was marginally consistent with the H.E.S.S. ULs.
    Given the numerous model parameters, it is possible that a different set of parameters can still allow a suitable fit of all data; nevertheless, our UL disfavors at least the parameter space discussed as fiducial by \citet{Zheng2024}.
    On the other hand, the VHE emission in the models by \citet{Ren2024} and \citet{Zhang2025} are below our UL.

   Next, considering our 4.1$\sigma$ excess as actual gamma rays from GRB~221009A, the VHE flux is quite consistent with the \citet{Zhang2025} model.
   While the VHE flux in the \citet{Ren2024} model falls somewhat short of our data, it may not be incompatible, considering the uncertainties in their best-fit parameters.
   It is notable that although the VHE fluxes at $\sim T^*+1\,$days in these two models are similar, their origins are quite different.
   For \citet{Ren2024}, the VHE emission at $\sim T^*+1$\,days is a continuation of what is seen by LHAASO, with the SSC emission from the inner jet dominating at all times.
   Contrastingly, for \citet{Zhang2025}, the SSC emission from the outer jet takes over as the dominant component starting from a few kiloseconds after $T^*$, as was also the case for \citet{Zheng2024}.
   Distinguishing between \citet{Ren2024} and \citet{Zhang2025} is not possible with the current data and requires more spectral and temporal coverage.

   Despite the apparent similarities in the model assumptions, comparison of the best-fit parameter values actually reveals significant differences among the three models.
   For example, the values of $n_{\rm CBM}(R=0)$ differ between the models by up to 4 orders of magnitude, and there are correspondingly large differences in the parameters related to the magnetic field $\epsilon_B$ and electron acceleration efficiency $\epsilon_\mathrm{e}$.
   Thus, even with the extensive MWL data obtained for GRB\,221009A, realistic structured jet models still appear to be underconstrained, and significant degeneracies in the parameters remain.
   An explicit comparison of the main model parameters for \citet{Ren2024}, \citet{Zheng2024}, and \citet{Zhang2025} is presented in Appendix~\ref{sect:model_params}.    

   Quantitatively evaluating the impact of our data for constraining the model parameters and the physics of structured jets is beyond the scope of this Letter.
   Below, we outline qualitatively how this may be approached with future high-sensitivity observations of GRBs with IACTs covering the earliest to latest possible timescales.
   For a given epoch (``snapshot''), a sufficiently well-measured broadband spectrum of the afterglow, particularly spectral breaks that reflect electron injection and cooling, and the SSC-to-synchrotron flux ratio mainly constrain the microphysical parameters such as the electron injection index $p$, $\epsilon_B$, and $\epsilon_\mathrm{e}$. 
   The VHE band is crucial as the emission is likely distinct from electron synchrotron radiation, with different dependences on the parameters \citep{Piran2010, Inoue_2013, Nava_2018}.
   In practice, this may not be straightforward as components from the inner and outer jets can overlap, but at least it can be done for the component dominating at that epoch.
   Obtaining a series of such snapshots throughout the afterglow evolution then constrains mainly the dynamical parameters $E_{\rm kin,0}(\theta)$, $\Gamma_{\rm b,0}(\theta)$, and $n_{\rm CBM}(R)$,
   leading us to valuable new information on the physics of GRB jet formation and propagation.
   There are also potential multimessenger implications for cosmic rays and neutrinos as discussed in \citet{Zhang2025}, whose model appears most consistent with our data. They attribute the emission exceeding 10\,TeV seen at $\sim T_0+[500,800]\,$s \citep{LHAASO2023KM2A} primarily to synchrotron radiation by ultrahigh-energy protons accelerated at the reverse shock, a hypothesis that could have been tested if sufficiently early follow-up by IACTs was possible.

\section{Conclusions}\label{sect:conclusions}

    GRB~221009A, the ``brightest-of-all-time'' GRB and first long GRB with strong evidence for a structured jet, was observed by LST-1.
    The high NSB in the hours following the GRB alert due to the full Moon prevented rapid follow-up by IACTs. 
    LST-1 was the first IACT to start observing at $T_0+1.33$\,days, covering epochs that were missed by other VHE facilities. Here we addressed for the first time the challenge of analyzing LST-1 data of a GRB in moonlight conditions, presenting a scheme adapted to handle the high-NSB conditions at the analysis level.
    We obtain an excess signal at a statistical significance of 4.1$\sigma$ for that night, constraining the intrinsic emission from GRB~221009A at the level of a few $10^{-11}\,\rm{erg\, cm^{-2}\, s^{-1}}$ at $E=0.3$--$5$\,TeV.
    ULs are derived at $T_0+3.33$\,days, still under moonlight conditions, and during later times, from $T_0+6.30$ to $T_0+17.32$\,days without the Moon. These analyses were cross-checked with an independent method and compatible results were obtained.
    Our data were compared to different realistic models of afterglows from structured jets that adequately describe published MWL data for GRB~221009A but imply significant differences in the VHE emission after about 1 day.
    Depending on the model, this can be dominated by SSC from either the narrow inner jet or the wider outer jet, and the flux can vary by more than an order of magnitude.
    Although all models were consistent with the LHAASO data and H.E.S.S. ULs, those that implied VHE flux at 1 day significantly exceeding $10^{-11}\,\rm{erg\, cm^{-2}\, s^{-1}}$ are disfavored by our results.
    If the observed excess corresponds to a gamma-ray signal from GRB~221009A, the VHE flux is consistent with a subset of the models, although we cannot distinguish between an inner and outer jet origin.
    Future sensitive IACT observations of GRBs over a broad range of timescales, together with comprehensive MWL data, should help to clarify the nature of structured jets and provide new insight into the physics of jets in long GRBs.

\section*{Author Contribution Statement}
    
    A. Aguasca-Cabot: LST-1 source-independent analysis, moonlight-adapted analysis study, project coordination, paper writing, and paper editing.
    P. Bordas: discussion of the LST-1 results and paper editing.
    A. Donini: principal investigator of the observations, project coordination, and paper drafting.
    S. Inoue: modeling discussion, paper writing, and paper editing.
    Y. Sato: discussion on modeling.
    M. Seglar~Arroyo: project coordination, paper writing, and paper editing.
    K. Terauchi: LST-1 source-dependent analysis, paper writing, and paper editing.
    K. Noda, S. Nozaki, J. Paiva, A. Tutone, and M. Will took care of the observations on 2022 October 10.
    The rest of the authors have contributed in one or several of the following ways: design, construction, maintenance, and operation of the instrument(s) used to acquire the data; preparation and/or evaluation of the observation proposals; data acquisition, processing, calibration, and/or reduction; production of analysis tools and/or related MC simulations; and discussion and approval of the contents of the draft.

\section*{acknowledgments}
    We sincerely thank Bing Theodore Zhang for providing us with detailed information about their theoretical model. We gratefully acknowledge financial support from the following agencies and organizations:

    Conselho Nacional de Desenvolvimento Cient\'{\i}fico e Tecnol\'{o}gico (CNPq), Funda\c{c}\~{a}o de Amparo \`{a} Pesquisa do Estado do Rio de Janeiro (FAPERJ), Funda\c{c}\~{a}o de Amparo \`{a} Pesquisa do Estado de S\~{a}o Paulo (FAPESP), Funda\c{c}\~{a}o de Apoio \`{a} Ci\^encia, Tecnologia e Inova\c{c}\~{a}o do Paran\'a - Funda\c{c}\~{a}o Arauc\'aria, Ministry of Science, Technology, Innovations and Communications (MCTIC), Brasil;
    Ministry of Education and Science, National RI Roadmap Project DO1-153/28.08.2018, Bulgaria;
    Croatian Science Foundation (HrZZ) Project IP-2022-10-4595, Rudjer Boskovic Institute, University of Osijek, University of Rijeka, University of Split, Faculty of Electrical Engineering, Mechanical Engineering and Naval Architecture, University of Zagreb, Faculty of Electrical Engineering and Computing, Croatia;
    Ministry of Education, Youth and Sports, MEYS  LM2023047, EU/MEYS CZ.02.1.01/0.0/0.0/16\_013/0001403, CZ.02.1.01/0.0/0.0/18\_046/0016007, \sloppy CZ.02.1.01/0.0/0.0/16\_019/0000754, CZ.02.01.01/00/22\_008/0004632 and CZ.02.01.01/00/23\_015/0008197 Czech Republic;
    CNRS-IN2P3, the French Programme d’investissements d’avenir and the Enigmass Labex, 
    This work has been done thanks to the facilities offered by the Univ. Savoie Mont Blanc - CNRS/IN2P3 MUST computing center, France;
    Max Planck Society, German Bundesministerium f{\"u}r Bildung und Forschung (Verbundforschung / ErUM), Deutsche Forschungsgemeinschaft (SFBs 876 and 1491), Germany;
    Istituto Nazionale di Astrofisica (INAF), Istituto Nazionale di Fisica Nucleare (INFN), Italian Ministry for University and Research (MUR), and the financial support from the European Union -- Next Generation EU under the project IR0000012 - CTA+ (CUP C53C22000430006), announcement N.3264 on 28/12/2021: ``Rafforzamento e creazione di IR nell’ambito del Piano Nazionale di Ripresa e Resilienza (PNRR)'';
    ICRR, University of Tokyo, JSPS, MEXT, Japan;
    JST SPRING - JPMJSP2108;
    Narodowe Centrum Nauki, grant number 2019/34/E/ST9/00224, Poland;
    The Spanish groups acknowledge the Spanish Ministry of Science and Innovation and the Spanish Research State Agency (AEI) through the government budget lines
    PGE2022/28.06.000X.711.04,
    28.06.000X.411.01 and 28.06.000X.711.04 of PGE 2023, 2024 and 2025,
    and grants PID2019-104114RB-C31,  PID2019-107847RB-C44, PID2019-104114RB-C32, PID2019-105510GB-C31, PID2019-104114RB-C33, PID2019-107847RB-C43, PID2019-107847RB-C42, PID2019-107988GB-C22, PID2021-124581OB-I00, PID2021-125331NB-I00, PID2022-136828NB-C41, PID2022-137810NB-C22, PID2022-138172NB-C41, PID2022-138172NB-C42, PID2022-138172NB-C43, PID2022-139117NB-C41, PID2022-139117NB-C42, PID2022-139117NB-C43, PID2022-139117NB-C44, PID2022-136828NB-C42, PDC2023-145839-I00 funded by the Spanish MCIN/AEI/10.13039/501100011033 and “and by ERDF/EU and NextGenerationEU PRTR; the "Centro de Excelencia Severo Ochoa" program through grants no. CEX2019-000920-S, CEX2020-001007-S, CEX2021-001131-S; the "Unidad de Excelencia Mar\'ia de Maeztu" program through grants no. CEX2019-000918-M, CEX2020-001058-M; the "Ram\'on y Cajal" program through grants RYC2021-032991-I  funded by MICIN/AEI/10.13039/501100011033 and the European Union “NextGenerationEU”/PRTR and RYC2020-028639-I; the "Juan de la Cierva-Incorporaci\'on" program through grant no. IJC2019-040315-I and "Juan de la Cierva-formaci\'on"' through grant JDC2022-049705-I. They also acknowledge the "Atracci\'on de Talento" program of Comunidad de Madrid through grant no. 2019-T2/TIC-12900; the project "Tecnolog\'ias avanzadas para la exploraci\'on del universo y sus componentes" (PR47/21 TAU), funded by Comunidad de Madrid, by the Recovery, Transformation and Resilience Plan from the Spanish State, and by NextGenerationEU from the European Union through the Recovery and Resilience Facility; “MAD4SPACE: Desarrollo de tecnolog\'ias habilitadoras para estudios del espacio en la Comunidad de Madrid" (TEC-2024/TEC-182) project funded by Comunidad de Madrid; the La Caixa Banking Foundation, grant no. LCF/BQ/PI21/11830030; Junta de Andaluc\'ia under Plan Complementario de I+D+I (Ref. AST22\_0001) and Plan Andaluz de Investigaci\'on, Desarrollo e Innovaci\'on as research group FQM-322; Project ref. AST22\_00001\_9 with funding from NextGenerationEU funds; the “Ministerio de Ciencia, Innovaci\'on y Universidades”  and its “Plan de Recuperaci\'on, Transformaci\'on y Resiliencia”; “Consejer\'ia de Universidad, Investigaci\'on e Innovaci\'on” of the regional government of Andaluc\'ia and “Consejo Superior de Investigaciones Cient\'ificas”, Grant CNS2023-144504 funded by MICIU/AEI/10.13039/501100011033 and by the European Union NextGenerationEU/PRTR,  the European Union's Recovery and Resilience Facility-Next Generation, in the framework of the General Invitation of the Spanish Government’s public business entity Red.es to participate in talent attraction and retention programmes within Investment 4 of Component 19 of the Recovery, Transformation and Resilience Plan; Junta de Andaluc\'{\i}a under Plan Complementario de I+D+I (Ref. AST22\_00001), Plan Andaluz de Investigaci\'on, Desarrollo e Innovación (Ref. FQM-322). ``Programa Operativo de Crecimiento Inteligente" FEDER 2014-2020 (Ref.~ESFRI-2017-IAC-12), Ministerio de Ciencia e Innovaci\'on, 15\% co-financed by Consejer\'ia de Econom\'ia, Industria, Comercio y Conocimiento del Gobierno de Canarias; the "CERCA" program and the grants 2021SGR00426 and 2021SGR00679, all funded by the Generalitat de Catalunya; and the European Union's NextGenerationEU (PRTR-C17.I1). This research used the computing and storage resources provided by the Port d’Informaci\'o Cient\'ifica (PIC) data center.
    State Secretariat for Education, Research and Innovation (SERI) and Swiss National Science Foundation (SNSF), Switzerland;
    The research leading to these results has received funding from the European Union's Seventh Framework Programme (FP7/2007-2013) under grant agreements No~262053 and No~317446;
    This project is receiving funding from the European Union's Horizon 2020 research and innovation programs under agreement No~676134;
    ESCAPE - The European Science Cluster of Astronomy \& Particle Physics ESFRI Research Infrastructures has received funding from the European Union’s Horizon 2020 research and innovation programme under Grant Agreement no. 824064.
    In addition, S. I. is supported by MEXT KAKENHI Grant Number 23H04896.

\bibliography{sample631}{}
\bibliographystyle{aasjournal}

\begin{appendix}   

    \section{Source-dependent analysis}
    \label{sect:srcdep-analysis}

    The results reported in this Letter have been validated with an independent analysis, known as ``source-dependent''.
    It is based on the a priori assumption that all gamma rays arrive from the same direction as in the case of a single pointlike gamma-ray source.
    A powerful parameter used in the source-dependent analysis is {\it dist}, the distance between the known source position and the centroid of the shower images. Since {\it dist} correlates with the shower impact parameter inside the light pool, it improves the energy reconstruction performance. Further details on the source-dependent analysis can be found in \cite{Abe_2023_LSTPerformance}.

    The same configurations as standard source-independent analysis are adapted for the MC production for both the moonlight and dark data sets.
    Specifically, the same techniques are used to match the NSB level of the simulation with that of the moonlight observations, and the same algorithms are used to optimize the charge extraction and the image-cleaning performance of LST-1 (see Section~\ref{sect:moon_analysis}).
    Parameters unique to the source-dependent analysis are introduced for the random forest generation.
    Differences in the analysis settings arise after the event reconstruction of each Cherenkov shower image.
    For the spectral analysis, the efficiency of the energy-dependent gammaness cut is 50\% and 80\% for the moonlight and dark data sets, respectively.
    The efficiency of the energy-dependent $\alpha$ parameter\footnote{The angle between the shower axis and the line between the known source position and the image centroid.} cut is 70\% for both the moonlight and dark data set.

\begin{figure}[h!]
        \centering
        \includegraphics[width=0.6\hsize]{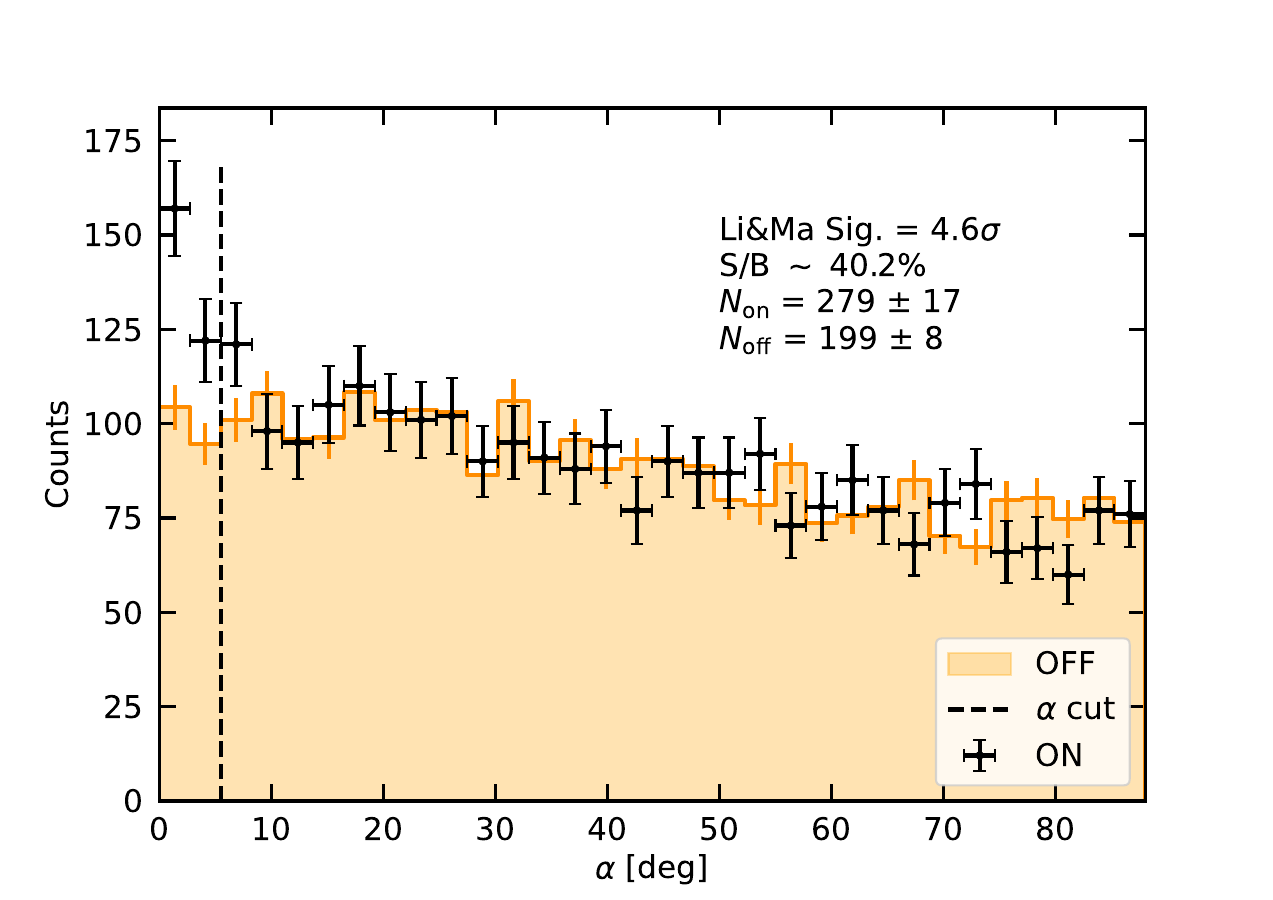}
        \caption{Same as Figure~\ref{fig:theta2plot_Oct10} but displaying the $\alpha$ distributions at $T_0+1.33$\,days after the analysis cuts. 
         The vertical dashed line indicates the $\alpha$ cut value, below which the detection statistical significance (Li$\&$Ma Sig.) is computed. The vertical error bars correspond to 1$\sigma$ statistical errors.}
        \label{fig:alphaplot_Oct10}
    \end{figure}

    The angular distributions of gamma events for the first night are shown in Figure~\ref{fig:alphaplot_Oct10}. Instead of the $\theta$ angle for the standard analysis, the $\alpha$ angle is used here to calculate the excess counts. 
    The applied cuts have been optimized from Crab observations during bright moonlight, which give similar NSB levels to the observations of GRB~221009A for the first days.
    A 4.6$\sigma$  statistical significance is obtained for the first night with excess counts of $80 \pm 22$, consistent with the source-independent results.
    We note that the signal-to-background ratio of the analysis is twice that of the source-independent analysis, as an expected consequence of the powerful background discrimination of the source-dependent analysis.

\begin{figure}[h!]
        \centering
        \includegraphics[width=1.0\hsize]{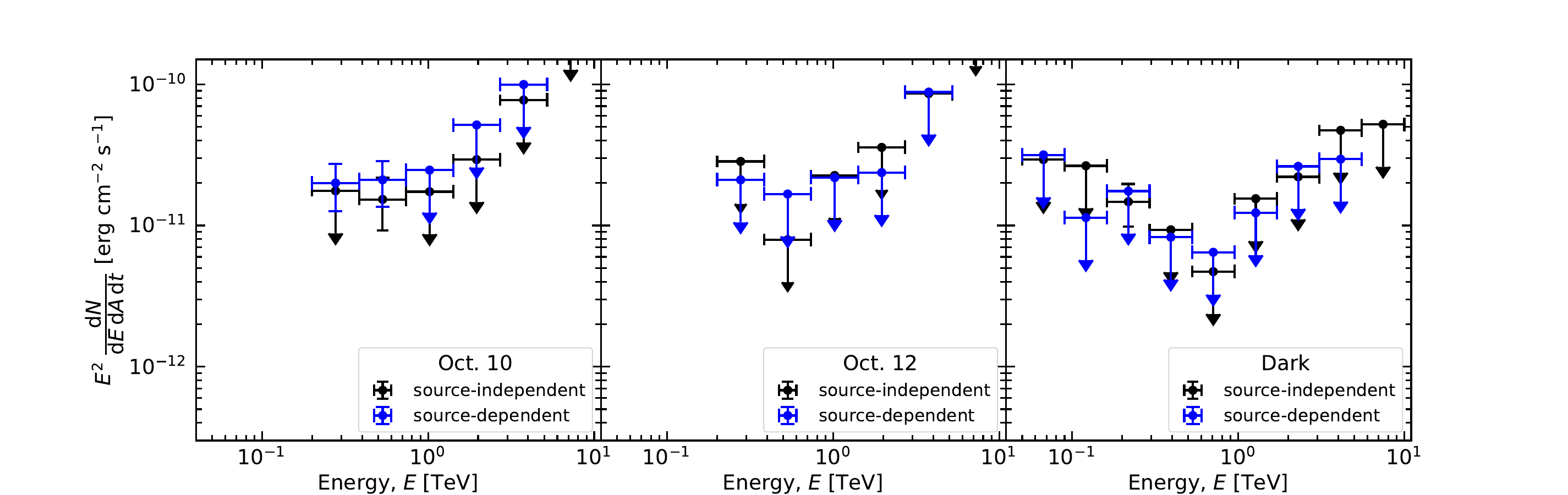}
        \caption{
        Comparison of SEDs between source-independent (black) and source-dependent (blue) approaches using the data sets on 2022 October 10 ($T_0+1.33\,$days; left), 2022 October 12 ($T_0+3.33\,$days; middle), and between 2022 October 15 and 27 (all dark, $T_0+[6.30,17.32]\,$days; right).
        ULs are computed at the 95\% confidence level when the TS is below 4. 
        Error uncertainties correspond to 1$\sigma$ statistical errors.
        }
        \label{fig:sed_comp}
    \end{figure}

\begin{figure}[h!]
        \centering
        \includegraphics[width=0.5\hsize]{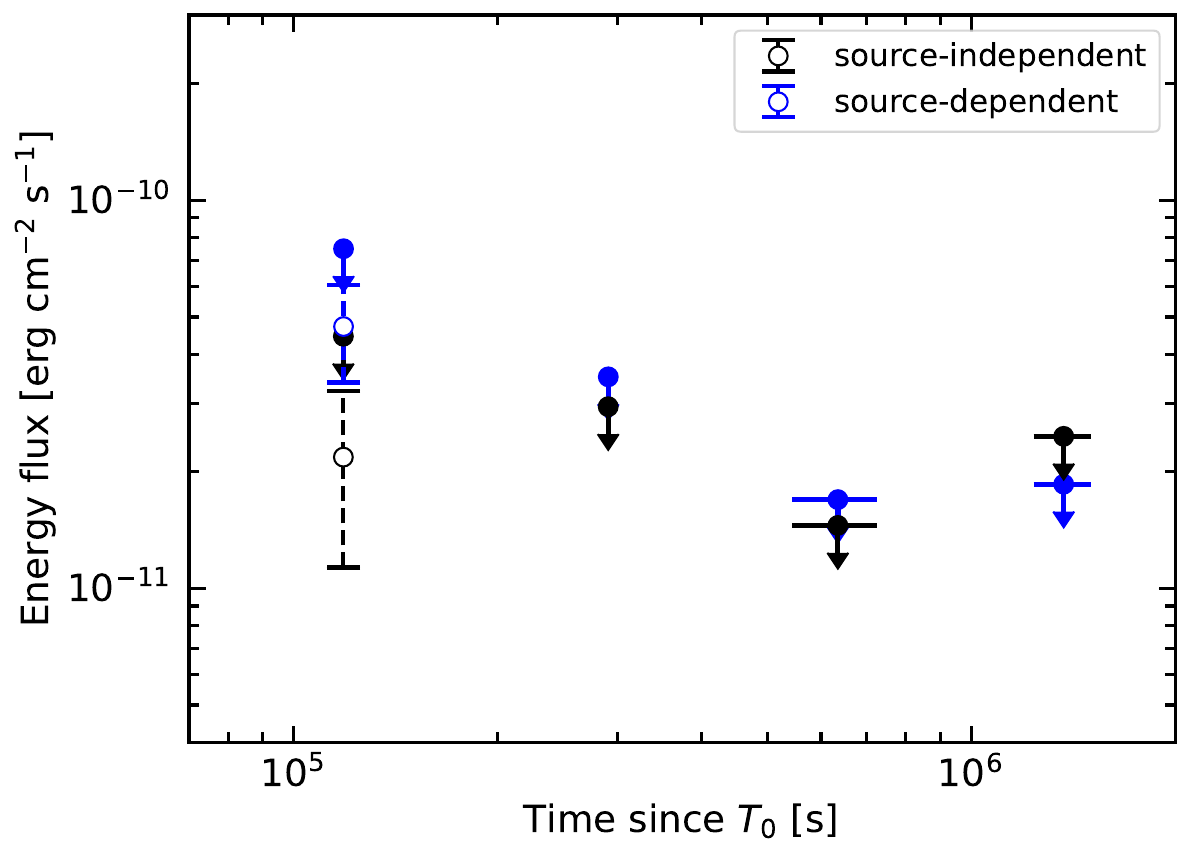}
        \caption{
        Comparison of intrinsic light curves (0.3--5\,TeV) between source-independent (black) and source-dependent (blue) approaches.
        As in Figure~\ref{fig:eflux-data-grb221009A}, both the energy flux and ULs at $T_0+1.33$\,days, are shown to illustrate the putative signal.
        Note that data points after $T_0+10^{6}$\,s cover different time ranges due to the different data selection criteria between the analyses.
        Error uncertainties correspond to 1$\sigma$ statistical errors, and ULs are computed at the 95\% confidence level.
        }
        \label{fig:lc_comp}
    \end{figure}

    The SED and light curve are computed with the 1D spectral analysis using \textit{gammapy}. Similarly to the source-independent analysis, the spectral model is defined as a simple power-law function with EBL attenuation (see Section~\ref{sect:results} for more details). The EBL-corrected SEDs computed with the source-dependent analysis are compared with the results of the source-independent analysis in Figure~\ref{fig:sed_comp}. We note that the ULs and SEDs are compatible, which serves as proof of the robustness of the results on GRB~221009A. It should be noted that on October 10, the SED point at $\sim$280 GeV ($\rm TS = 7.8$) is obtained for the source-dependent analysis, while a UL is obtained for the source-independent analysis.
    The fact that the source-dependent analysis resulted in a slightly higher detection significance may support this result, although possible systematics between the analyses cannot be ruled out, given the systematic positive bias in the flux observed on October 10.
    The comparison of the intrinsic light curve in terms of integral energy flux is shown in Figure~\ref{fig:lc_comp}.
    As in Figure~\ref{fig:eflux-data-grb221009A}, both the energy flux and ULs at $T_0+1.33$\,days are shown to illustrate the putative signal. 
    The energy flux points at $T_0+1.33$\,days for both analyses are compatible at almost 1$\sigma$ statistical error. This effect can be explained by systematic uncertainties and the difference between the two analysis methods, as indicated in \citet{Abe_2023_LSTPerformance}.

\section{Comparison of model parameters}
    \label{sect:model_params}
    Here we present a brief graphic comparison of the parameters in the structured jet afterglow models of \citet{Ren2024}, \citet{Zheng2024}, and \citet{Zhang2025}.
    The aim is to simply illustrate the sizable disparities in the best-fit parameters between the models, rather than a detailed discussion of the physical implications or the quality of fits to the observed data, which is beyond our scope.

\begin{figure}[h!]
        \centering
        \begin{minipage}{0.48\textwidth}
            \centering
            \includegraphics[width=\textwidth]{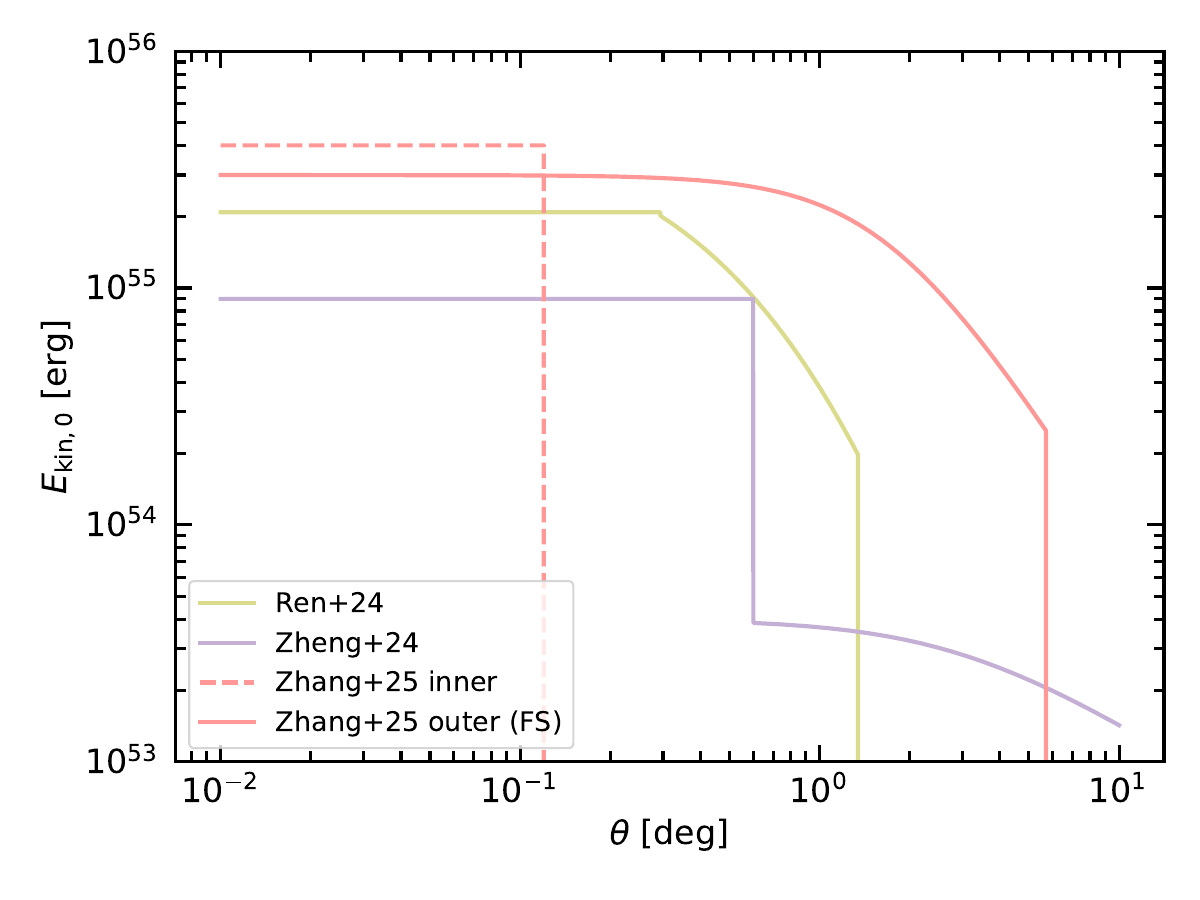}
        \end{minipage}
        \begin{minipage}{0.48\textwidth}
            \centering
            \includegraphics[width=\textwidth]{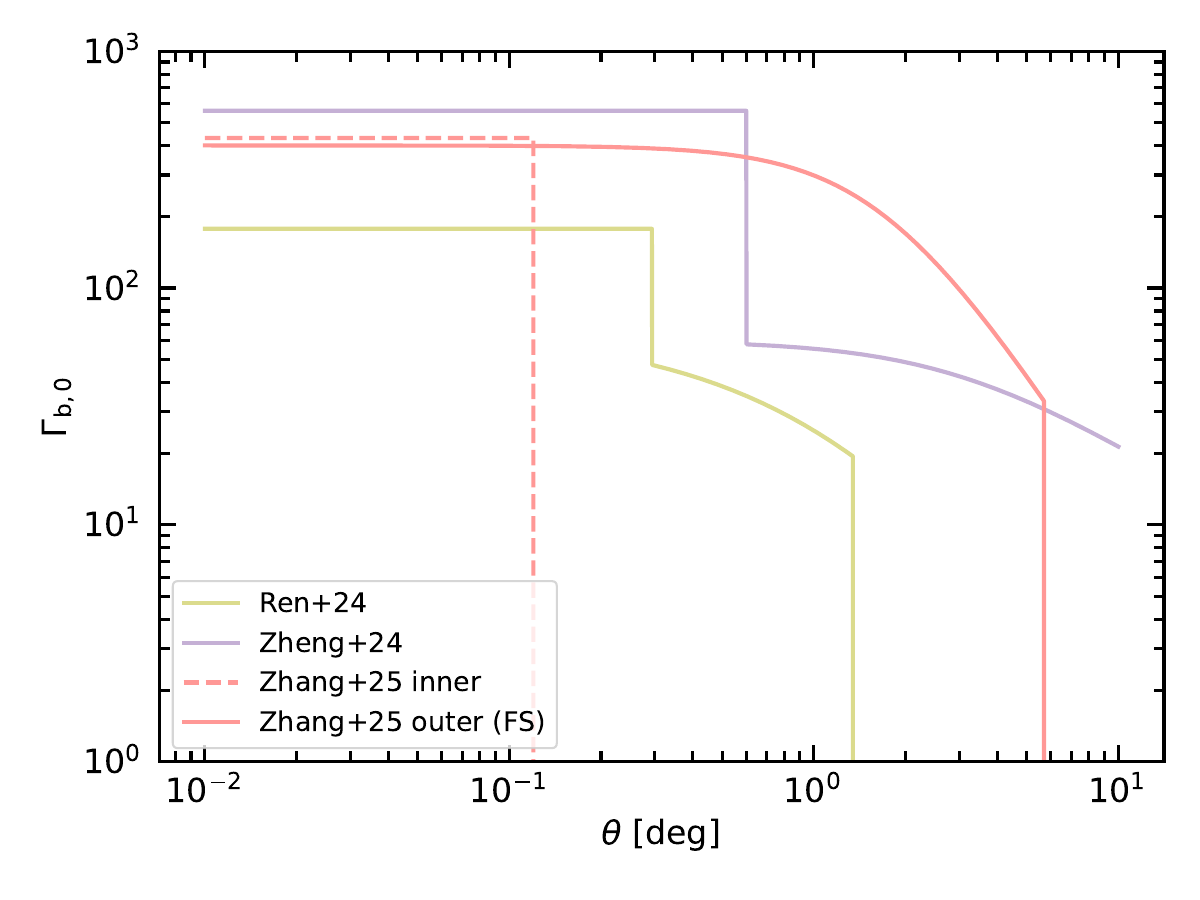}
        \end{minipage}

        \vspace{0.5cm}
        
        \begin{minipage}{0.48\textwidth}
            \centering
            \includegraphics[width=\textwidth]{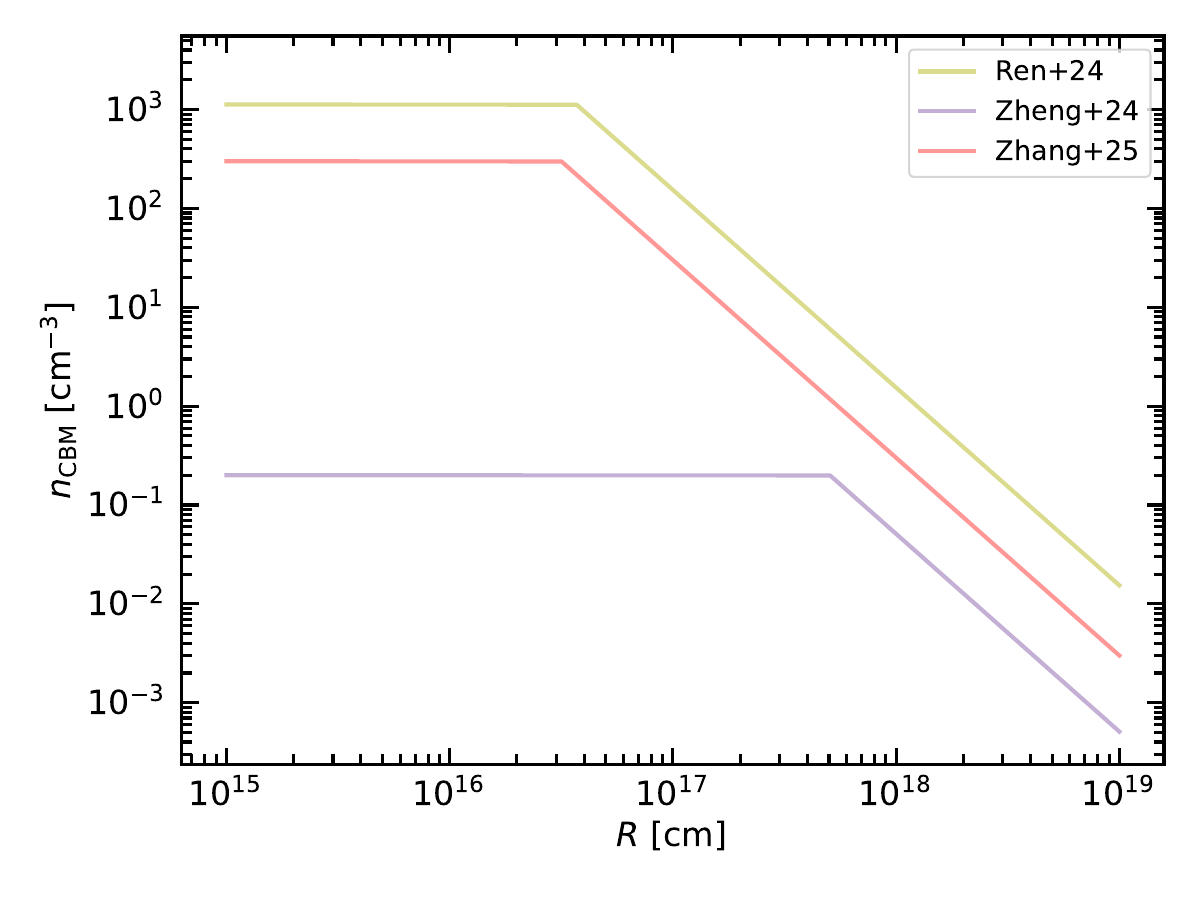}
        \end{minipage}        
        \begin{minipage}{0.48\textwidth}
            \centering
            \includegraphics[width=\textwidth]{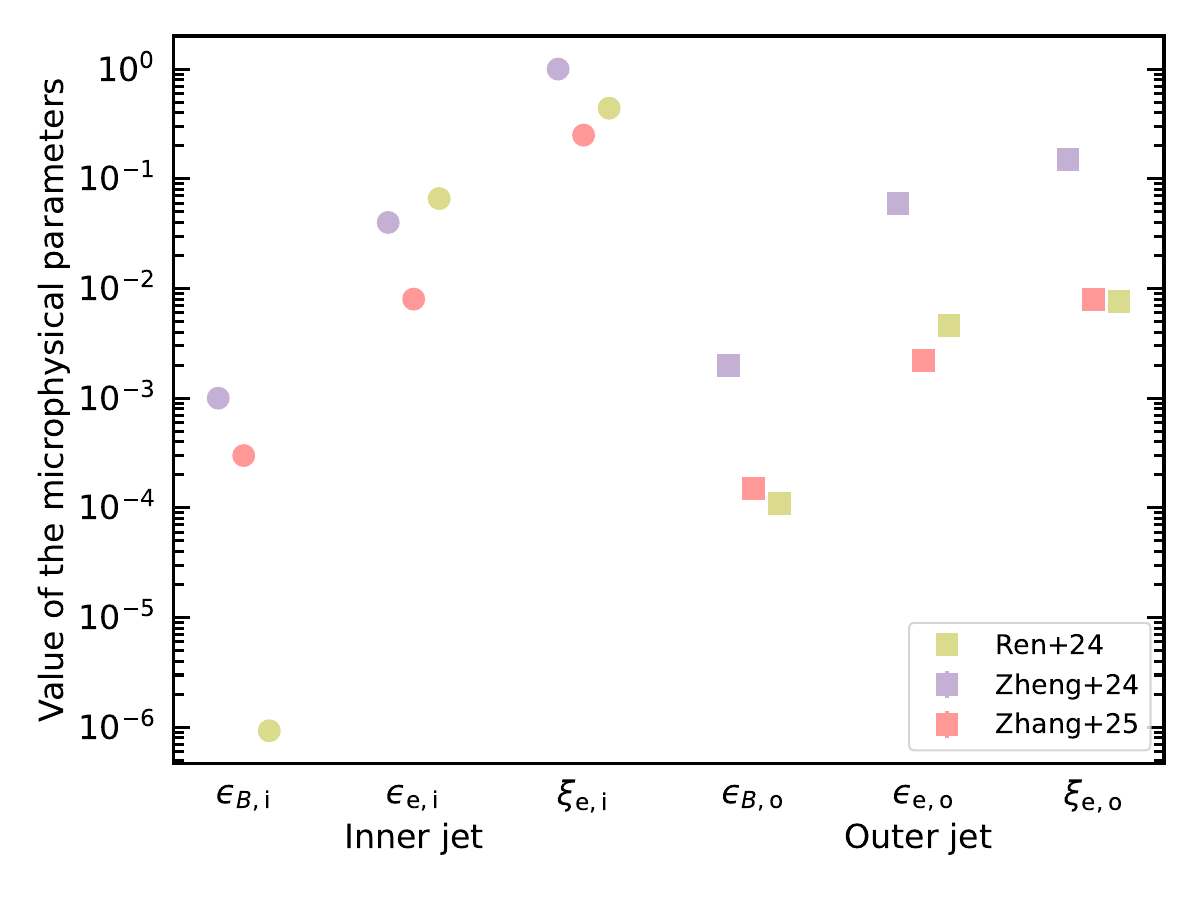}
        \end{minipage}
        
        \caption{Comparison of model parameters for \citet{Ren2024}, \citet{Zheng2024}, and \citet{Zhang2025}.
        Top left: $E_{\rm kin,0}(\theta)$, initial kinetic energy vs. angle from jet axis.
        Top right: $\Gamma_{\rm b, 0}(\theta)$, initial bulk Lorentz factor vs. angle from jet axis.
        Bottom left: $n_{\rm CBM}(R)$, circumburst medium density vs. radius.
        Bottom right:
        $\epsilon_{B,\rm{i}}$, fraction of postshock energy in the magnetic field; $\epsilon_{\rm e,i}$, fraction of postshock energy in accelerated electrons; $\xi_{\rm e,i}$, fraction in number of accelerated electrons for the inner jet;
        and $\epsilon_{B,\rm{o}}$, $\epsilon_{\rm e,o}$ and $\xi_{\rm e,o}$, analogous quantities for the outer jet.
        }
        \label{fig:model_parameters}
    \end{figure}

    Figure~\ref{fig:model_parameters} 
    compares $E_{\rm kin,0}(\theta)$ (initial kinetic energy versus angle from jet axis),
    $\Gamma_{\rm b, 0}(\theta)$ (initial bulk Lorentz factor versus angle from jet axis),
    $n_{\rm CBM}(R)$ (circumburst medium density versus radius),
    and the microphysical parameters $\epsilon_{B,\rm{i}}$ (fraction of postshock energy in the magnetic field), $\epsilon_{\rm e,i}$ (fraction of postshock energy in accelerated electrons), and $\xi_{\rm e,i}$ (fraction in number of accelerated electrons) for the inner jet, as well as the analogous parameters $\epsilon_{B,\rm{o}}$, $\epsilon_{\rm e,o}$ and $\xi_{\rm e,o}$ for the outer jet.
    Only \citet{Ren2024} provide errors for the parameters from a Markov Chain MC analysis, and these are not shown for consistency in the comparison.

    Particularly large differences between the models can be seen for $\epsilon_{B,\rm{i}}$ and $n_{\rm CBM}$ at small $R$ and $E_{\rm kin,0}(\theta)$ and $\Gamma_{\rm b, 0}(\theta)$ at large $\theta$.
    This indicates that for constraining such structured jet afterglow models with large number of parameters, the available MWL data are insufficient and require more extensive spectral and temporal coverage including VHE observations.
        
\end{appendix}




\end{document}